\documentclass[12pt,preprint]{aastex}
\renewcommand{\cite}{\citealp}

\begin{document}

\title{Theoretical fits of the $\delta$ Cephei light, radius and radial velocity curves}

\author{Giovanni Natale\altaffilmark{1,2}, Marcella Marconi\altaffilmark{2}, Giuseppe Bono\altaffilmark{3}}  

\altaffiltext{1}{Max-Planck-Institut fuer Kernphysik Saupfercheckweg 1, D-69117 Heidelberg (current position), giovanni.natale@mpi-hd.mpg.de}
\altaffiltext{2}{INAF-Osservatorio Astronomico di Capodimonte, Via Moiariello 16, 80131 Napoli, Italy; marcella@oacn.inaf.it}
\altaffiltext{3}{INAF-Osservatorio Astronomico di Roma, Via Frascati 33, 00040 Monte Porzio Catone, Italy; bono@mporzio.astro.it}

\begin{abstract}
We present a theoretical investigation of the light, radius and 
radial velocity variations of the prototype $\delta$ Cephei. We find that 
the best fit model accounts for luminosity and velocity amplitudes with an 
accuracy better than $0.8\sigma$, and for the radius amplitude with an 
accuracy of $1.7\sigma$. The chemical composition of this model suggests 
a decrease in both helium (0.26 vs 0.28) and metal (0.01 vs 0.02) 
content in the 
solar neighborhood. Moreover, distance determinations based on the fit 
of light curves agree at the $0.8\sigma$ level with the trigonometric 
parallax measured by the Hubble Space Telescope (HST). On the other hand, 
distance determinations based on angular diameter variations, that are
independent of interstellar extinction and of the $p$-factor 
value, indicate an increase of the order of 5\% in the HST parallax.  
\end{abstract}


\section{Introduction} 

Classical Cepheids are the most important primary distance indicators. One of 
the key issues concerning the use of Cepheids as distance 
indicators is their dependence on chemical composition. Empirical tests 
seem to suggest that metal-rich Cepheids are, at fixed period, brighter than 
metal-poor ones, either over the entire period range \citep[][]{KSS98,KSF03,K03,SCG04,G04,S04,M06} or for periods shorter than $\approx 25$ days  \citep[][]{SBR97,Sa04}. 
On the other hand, distance estimates based on the Near-Infrared Surface
Brightness method indicate that the slope of the Period-Luminosity ($PL$)
relation might be the same for Magellanic and Galactic Cepheids \citep{GSB05}.
However, the impact of the projection factor $p$ on the Cepheid distances is still debated \citep[][]{S06,N07}.
Early theoretical predictions based on linear pulsation models 
\citep[e.g.][]{SBT99,A99,BA01} predicted a mild metallicity effect, 
while nonlinear convective models 
\citep[][]{BMS99,C02,F02,MMF05} showed that the  metallicity  
affects both the zero-point and the slope of the $PL$ relations. 
This effect is more evident in the optical bands and 
it is nonlinear when moving from metal-poor ($Z=0.004$) to super 
metal-rich ($Z=0.04$) structures. Current predictions 
also suggest a turnover across solar metallicity ($Z$ $\sim$0.02) due 
to the correlated increase in He content \citep[][]{F02,MMF05}. 

Three approaches are adopted to calibrate 
the Cepheid PL relations: cluster distances, trigonometric parallaxes 
and Baade-Wesselink (BW) method. Parallaxes based on Cepheids in open 
clusters and associations have been widely adopted to calibrate the PL relations \citep{TB02}, but \citet{Fo07} found a systematic difference with other methods. Accurate Cepheid parallaxes
have been provided by \citet{BFM02} and by \citet{BFM07} using 
the Fine Guide Sensor on board the Hubble Space Telescope (HST). 
They measured the distance of several Galactic Cepheids with 
average relative errors of $\approx 8$\%.  
The BW method appears to be a robust approach, but 
some assumptions need to be investigated 
\citep[][]{G87,BCS94,GSB05}. The value of the projection 
factor ($p$), i.e. the parameter adopted to 
transform radial into pulsation velocity, is still 
controversial \citep[][]{S95,MK062}. The 
typical value used in the 
literature is $p=1.36$ \citep{BMB82}, but period-dependent values 
are also adopted \citep[][]{GFG93,GSB05}. 
However, the $p$-factor of $\delta$ Cephei, $p=1.27\pm 0.06$, measured 
by \citet{MK05}, using the HST parallax, agrees quite well with the  
predicted value \citep{NFM04}. 

Although, several Cepheids are binaries dynamical mass estimates are 
only available for a handful objects \citep[][]{E05}. 
Current mass estimates rely either on evolutionary or on pulsation 
prescriptions \citep[][]{DC00,KW02,CBFMM05} and we are still facing 
the long-standing problem of the ``Cepheid mass discrepancy'' \citep[]{Cox80}. 
Evolutionary masses are systematically larger than pulsation masses 
\citep[][]{B01,BBK01,CBFMM05}.

Theoretical fits of Cepheid light curves provide robust constraints 
on their intrinsic parameters and distances \citep[][]{WAS97,BCM02,KW06}. 
This approach, up to now, has only been applied to Bump Cepheids.  
We present, for the first time, a simultaneous 
fit of optical/NIR light, radial velocity, and angular diameter 
variations of the prototype $\delta$ Cephei. We selected this
object because accurate measurements of light, radius, velocity 
curves, and of the $p$-factor are available and its geometric 
distance is known with an accuracy of 4\% \citep[273$^{+12}_{-11}$][]{BFM02}.

\section{Empirical and theoretical framework}

The optical ($V,I$) and the near-infrared (NIR) $K$-band light curve were 
collected by \citet{K98} and by \citet{BFFN97}.
We selected these light curves because both the shape and the amplitude in the 
optical bands depend on surface temperature and radius variation, while in 
$K$-band the dependence on temperature is significantly reduced. The radial 
velocity data were collected by 
\citet{BBM94}\footnote{See the database 
http://crocus.physics.mcmaster.ca/Cepheid/}, while the angular diameter and the projection 
factor by \citet{MK05}. The observables for $\delta$ Cephei are summarized in 
Table~1.  

The adopted theoretical framework is based on a nonlinear radial pulsation code,
including the nonlocal and time-dependent treatment of turbulent convection 
(\citep{S82,BS94,BMS99}. The system of nonlinear 
equations is closed using a free parameter, $\alpha_{ml}$, that 
is proportional to the mixing length parameter. Changes in the $\alpha_{ml}$ 
parameter affect, as expected, both the limit cycle stability (pulsation 
amplitudes) and the topology of the instability strip \citep{F07,DMC04}. 
Similar approaches for the treatment of convective transport  have been 
developed by \citet{Feu99,BS07}, and \citet{OW05}.   

In order to provide a detailed fit of $\delta$ Cephei pulsation properties 
we adopted three different metal abundances, ranging from $Z=0.01$ to 
$Z=0.02$, to account for current spectroscopic measurements \citep{FC97} 
that indicate a metal abundance close to solar ($[Fe/H]=-0.01\pm0.06$). 
Recent spectroscopic measurements of solar heavy element abundances show 
a significant decrease when compared with previous estimates, and indeed 
the ratio between heavy-elements and hydrogen, $Z/X$, decreased from 0.023 
\citep{GS98} to 0.0177 \citep{L03} or to 0.0171 \citep{A04,A05}. 
This implies that the surface heavy element abundance of the Sun decreased 
from $Z=0.017$ to $Z=0.0126/0.0133$. 
The surface He abundance of the Sun 
according to helioseismological estimates, combined with these latest solar 
metallicity estimates, is $Y=0.2485\pm0.0034$ \citep{BA04}. However, 
the new solar photospheric abundances are at odds with helioseismology 
measurements \citep{Bah05,Gu05} and, at present, solar models are constructed by adopting protosolar helium 
abundances ranging from $Y_0\sim0.26$  to $Y_0\sim0.28$ and heavy metal 
abundances ranging from $Z_0=0.015$ to $Z_0=0.02$. According to these 
evidence we adopted $Z=0.02$ and $Y=0.28$ (see Table~2). Once the chemical 
composition was fixed, we constructed a sequence of isoperiodic pulsation 
models, by adopting the pulsation relation 
(the relation connecting the period to stellar mass, luminosity, 
and effective temperature) provided by \citet{BCM00}. We selected 
the fundamental models with periods within 2\% of the observed period 
(P=5.3663 days, 
Fernie 1995\footnote{http://www.astro.utoronto.ca/DDO/research/cepheids/}) 
and pulsationally unstable according to the analytical 
relations for the instability strip boundaries provided by \citet{BCM00}. 
For each mass value, we fixed the lower limit to the input luminosity 
according to the  Mass-Luminosity (ML) relation given by \citet{BCCM00} 
and the upper limit 
according to mild overshooting predictions, i.e. by adding to the quoted  
value a fixed luminosity excess i.e. $\Delta logL=0.25$ \citep[][]{CWC93}. 
The bolometric light curves were transformed into the UBVIJK 
photometric bands by adopting the model atmospheres provided by 
\citet{CGK97}. 

\section{Results and discussion}
We constructed a sequence of models at fixed chemical composition 
($Z=0.02$, $Y=0.28$) and by assuming steps in mass and in effective 
temperature of $\Delta M = 0.5M_\odot$ and $\Delta T_e=100K$. 
The comparison between theory and observations indicate that a model 
with $M=5.5M_\odot$, and $Te=5700K$ accounts for the shape of the 
light curves and for the secondary bump located on the decreasing branch 
of the radial velocity curve ($\phi \approx 0.2$). To further improve the 
agreement between predicted and empirical observables, we constructed 
a new sequence of models by adopting smaller steps in mass and in effective 
temperature, namely $\Delta M=0.1M_\odot$,$\Delta T_e=25 K$. For 
every model in this sequence, the mixing length parameter was changed 
from $\alpha_{ml}=0.3$ to $\alpha_{ml}=0.4$. Note that an increase in  
$\alpha_{ml}$ causes a decrease in the pulsation driving, and in turn 
a decrease in the limit cycle amplitudes.
Data plotted in the left panels of Fig. 1 show that the best fit model
($\alpha_{ml}=0.33$) accounts for the optical and NIR light curves, but 
both the radial velocity ($\sigma(\Delta V_r)=4.2$) and the radius 
($\sigma(\Delta R)=3.3$) amplitudes are systematically smaller than 
the observed ones. This discrepancy can only be removed by decreasing 
the $p$-factor from 1.27 to the barely plausible value of $p \sim 1$. 
To further constrain the impact of the adopted chemical composition 
on the limit cycle stability, we constructed a new sequence of models 
by using the same metal abundance and a lower He content ($Y=0.26$). 
The new best fit model (see Table~2 and the middle panels 
of Fig. 1) accounts for optical and NIR light curves and better 
reproduces radial velocity ($\sigma(\Delta V_r)=1.6$) and radius 
($\sigma(\Delta R)=2.3$) variations. The predicted $V$-band light 
curve presents a mild discrepancy along the rising branch. However, 
the discrepancy on luminosity amplitudes is smaller than $0.4\sigma$.  
In order to verify the occurrence of a degeneracy in the input 
parameters, we also computed a new series of models by using a lower 
He content, ($Y=0.24$, see Table~2 and right panels of Fig. 1). 
Interestingly enough, velocity and radius amplitudes predicted by 
new best fit-model are at least $2.5\sigma$ smaller than observed. 
Once again, a reasonable fit between theory and observations would 
require a barely plausible $p$-factor ($p=1.03$). 

In order to test how the fits depend on metallicity, 
we computed a new series of models by adopting $Z=0.015$, and He content
of $Y=0.27$ (see Table~2). Data plotted in the left panels of Fig. 2 show 
that the new best fit model agrees reasonably well with the observations. 
However, the discrepancy in the radial velocity is $1.6\sigma$ and in 
the radius amplitude is $2.5\sigma$. This finding indicates that a 
decrease in the metal content improves the agreement between theory 
and observations. Therefore, we constructed 
two new sequences of pulsation models by adopting a more metal-poor 
chemical composition ($Z=0.01$) and two different He contents 
($Y=0.26, 0.24$). The new best fit models (see Table~2, middle 
and right panels of Fig. 2) account quite well for luminosity, 
radial velocity, and radius variations. A marginal difference 
between theory and observations is still present. However,  
the discrepancy is on average smaller than $1\sigma$ for $Y=0.26$ 
and systematically larger for $Y=0.24$ (see Table~2).  
The comparison between theory and observations brings forward several findings. 
{\em i)} the more metal-poor ($Z=0.01$) best fit models account for the 
observed light and velocity amplitudes at a level better than $0.8\sigma$ 
and the average discrepancy over the entire cycle is $< 0.2\Sigma$ 
(see Table~2). 
{\em ii)} The fit of the radial velocity curve is very sensitive to the 
input parameters. The small ``bump`` located along the decreasing branch
($\phi=0.2$) mainly depends on the mass value, and it is independent of 
the chemical composition. Moreover, the radial velocity amplitude 
strongly depends on the He content. For example, an increase/decrease 
of $\sim 8$\% in He content causes, at Z=0.02, an increase on the amplitude 
discrepancy of at least $1.4\sigma$. The more metal-rich best fit models 
(Y=0.28, 0.25) to properly fit the velocity amplitude would require 
$p$-factor values that are at least $\sim 20$\% smaller than the 
empirical measurement by \citet{MK062}. Therefore, they do not pass 
the empirical validation.  
{\em iii)} Predicted radius variations are at least $1.7\sigma$ smaller than 
observed while the average discrepancy over the entire curve is never 
smaller than $0.2\Sigma$ for all the adopted chemical compositions. 
The mismatch is mainly caused by the few points located across the phases 
of minimum radius. We do not have a firm explanation for this discrepancy. 
These phases are the most rapid ones along the pulsation cycle and the radial 
displacements of the outermost layers might be affected either by shock waves 
or by velocity gradients \citep{Ma06}. 
This critical point deserves further investigations, since radius measurements 
depend on the adopted distance and on limb darkening correction, but they  
are affected neither by the interstellar extinction nor by the $p$-factor. 
Therefore, they can play a key role in constraining theoretical predictions. 
{\em iv)}  The model with $Z = 0.01$ and $Y = 0.26$ provides the optimum fit to all 
observables (except for the only three magnitudes on the rising branch of 
    the V light curve). This finding taken at face 
value indicates that the He content might be 8\% smaller than typically 
assumed to construct Galactic Cepheid 
models. The metal content of $\delta$ Cephei seems to be $0.1-0.2$ dex more 
metal-poor than suggested by spectroscopy, but this constraint is less robust. 
Note that the $p$-factor adopted for this model ($p=1.28$) 
agrees quite well with the measurement by M\'erand et al. (2006). 
{\em v)} The best fit models present a luminosity excess (see column 7 
in Table~2) that ranges from $\Delta Log L/L_\odot=0.07$ (Z=0.01, Y=0.26) 
to 0.17 (Z=0.02, Y=0.24). These values are between the luminosities predicted 
by a canonical ML relation \citep{BCCM00} and those predicted by mild core 
overshooting ML relation ($\Delta Log L=0.25$, Chiosi et al. 1993). 

As a further validation of the current approach, we also estimated the absolute 
distance to $\delta$ Cephei using the best fit models that passed the empirical 
validations. In order to provide a plausible 
estimate of the errors affecting distance determinations we assumed the largest 
uncertainties on the adopted input parameters:  
$\Delta M=0.1M_\odot$, $\Delta T=50 K$, $\Delta LogL=0.02$, 
$\Delta R=0.5R_\odot$. 
By accounting for the entire error budget the intrinsic uncertainties 
on current distance determinations is about $3\%$ for photometric distances 
and $\approx 5$ pc for the distances derived from the fit of radius variations. 
Distance estimates based on the apparent distance moduli provided by 
the fit of both optical and NIR light curves and on the reddening law provided  
by \citet{CCM89} are listed in columns 6 and 7 of Table~3. 
The comparison with the HST parallax shows that distances based on $V,I$-bands
agree at the $0.2\sigma$ level, while those based on the $V,K$-bands are 
within $0.8\sigma$ level. 
Errors either in the photometric zero-points 
or in the adopted color-temperature transformations or in the extinction 
law account for the difference in the distance determinations based on light 
curves. Moreover, the occurrence of a circumstellar envelope (CSE) around 
$\delta$ Cephei \citep{MK062} could also affect the reddening corrections 
due to the different relative distribution of the absorbing grains around 
the star, compared to typical interstellar medium distribution.
Independent distance determinations follow from comparing the observed angular diameter variations by \citet{MK05} with the radius curves from four different models. The distances (Table~3, column 8) agree well with each other and are on 
average only 5\% ($1.3 \sigma$) 
larger than the trigonometric distance. The 
comparison of observed and model radii is in principle a powerful method
to derive Cepheid distances, since it is independent of the assumption 
on the $p$-factor and on interstellar extinction.   
Note that an increase in the distance would also imply an increase in the 
$p$-factor, since at fixed radial velocity amplitude, 
$p\propto \Delta \Theta\cdot d$, where $\Delta \Theta$ is the angular diameter amplitude.

Finally we mention that the mean radii predicted by the best fit 
models match within $1\sigma$ the empirical estimate based on the 
angular diameter provided by ME06 ($R=43\pm2 R_\odot$), by assuming 
the HST parallax.

\acknowledgments
It is a pleasure to thank A. M\'erand for several interesting discussions 
on interferometric measurements. We also acknowledge an anonymous referee
for a detailed and helpful report.

{}

\clearpage
\begin{center} 
\begin{deluxetable}{lllll lllll l}
\tablewidth{0pt}
\tabletypesize{\scriptsize}
\tablecaption{Direct and derived observables of $\delta$ Cephei.  
}\label{tbl-1}
\tablehead{
\colhead{$m_V$\tablenotemark{a}}&
\colhead{$m_I$\tablenotemark{a}}&
\colhead{$m_K$\tablenotemark{a}}&
\colhead{$A_V$\tablenotemark{a}}&
\colhead{$A_I$\tablenotemark{a}}&
\colhead{$A_K$\tablenotemark{a}}&
\colhead{$\Delta V_r$\tablenotemark{a}}&
\colhead{R\tablenotemark{a}}&
\colhead{$\Delta R$\tablenotemark{a}}&
\colhead{$\mu$\tablenotemark{a}}&
\colhead{$p$\tablenotemark{a}}
}
\startdata
 4.02 & 3.23 & 2.34 & 0.84  & 0.53  & 0.20  & 47.8& $43\pm2$ &$5.4\pm0.2$ & $274\pm11$ & $1.27\pm0.06$     
\enddata
\tablenotetext{a}{Apparent $V,I,K$-band mean magnitudes and luminosity amplitudes \citep{K98,BFFN97}, radial velocity amplitude \citep[km/sec,][]{BBM94},  mean radius \citep[solar units,][]{MK062}, 
radius amplitude \citep[][]{MK062}, true distance \citep[pc,][]{BFM02}, 
$p$-factor \citep{MK062}. 
}
\end{deluxetable}
\end{center} 

\begin{center} 
\begin{deluxetable}{lllll lllll lllll llll}
\tablewidth{0pt}
\tabletypesize{\scriptsize}
\tablecaption{Adopted intrinsic parameters of the best fit models and discrepancies 
between theory and observations.  
}\label{tbl-2}
\tablehead{
\colhead{Z\tablenotemark{a}}&
\colhead{Y\tablenotemark{a}}&
\colhead{M\tablenotemark{a}}&
\colhead{L\tablenotemark{a}}&
\colhead{$T_e$\tablenotemark{a}}&
\colhead{R\tablenotemark{a}}&
\colhead{$\Delta L$\tablenotemark{a}}&
\colhead{$\alpha$\tablenotemark{a}}&
\colhead{$p$\tablenotemark{a}}&
\colhead{$\Sigma_V$\tablenotemark{b}}&
\colhead{$\Sigma_I$\tablenotemark{b}}&
\colhead{$\Sigma_K$\tablenotemark{b}}&
\colhead{$\Sigma_{Vr}$\tablenotemark{b}}&
\colhead{$\Sigma_R$\tablenotemark{b}} &
\colhead{$\sigma_V$\tablenotemark{c}}&
\colhead{$\sigma_I$\tablenotemark{c}}&
\colhead{$\sigma_K$\tablenotemark{c}}&
\colhead{$\sigma_{Vr}$\tablenotemark{c}}&
\colhead{$\sigma_R$\tablenotemark{c}} 
}
\startdata
0.02&0.28&  5.5&3.31&5800&45.06&0.11& 0.33&0.94 & 0.04  & 0.03  & 0.01  & 0.4  &0.3 &0.9& 0.3& 0.3& 4.2 & 3.3    \\
0.02&0.26&  5.5&3.32&5800&45.07&0.15& 0.40&1.15 & 0.2  & 0.07  & 0.008  & 0.1  &0.2 &0.2& 0.4& 0.3& 1.6 & 2.3    \\
0.02&0.24&  5.5&3.29&5700&45.32&0.17& 0.38&1.03 & 0.08  &0.05   &0.01   & 0.2  &0.3 &0.1& 0.5& 0.0& 3.0 & 2.8    \\
0.015&0.27& 5.5&3.31&5750&45.29&0.08& 0.40&1.18 & 0.2  & 0.06  &0.009   & 0.1  &0.2 &0.5& 0.7& 0.3& 1.6 & 2.1    \\
0.01&0.26&  5.5&3.33&5800&46.07&0.07& 0.43&1.28 & 0.2  & 0.07  &0.01   & 0.07  &0.2 &0.6& 0.8& 0.5& 0.4 & 1.7    \\
0.01&0.24&  5.5&3.33&5800&46.08&0.11& 0.44&1.21 & 0.1  & 0.05  &0.009   &0.1   &0.2 &0.5& 0.7& 0.5& 1.0 & 2.0    
\enddata
\tablenotetext{a}{Metal ($Z$) and He ($Y$) abundance by mass, stellar mass ($M/M_\odot$), 
logarithmic luminosity ($\log L/L\odot$), effective temperature (K), radius ($R/R_\odot$), luminosity excess (dex) 
respect to canonical ML relation from \citet{BCCM00}, convection efficiency parameter ($\alpha_{ml}$), projection factor required to fit the observed amplitudes.}
\tablenotetext{b}{Standard deviations of the fit between the predicted curves and 
observations over the entire cycle.} 
\tablenotetext{c}{Standard deviations between observed and predicted amplitudes.}
\end{deluxetable}
\end{center}

\begin{center} 
\begin{deluxetable}{lllllllllll}
\tablewidth{0pt}
\tabletypesize{\scriptsize}
\tablecaption{Distance and absolute magnitudine determinations based on different methods.  
}\label{tbl-2}
\tablehead{
\colhead{Z\tablenotemark{a}}&
\colhead{Y\tablenotemark{a}}&
\colhead{$\mu_V$\tablenotemark{a}}&
\colhead{$\mu_I$\tablenotemark{a}}&
\colhead{$\mu_K$\tablenotemark{a}}&
\colhead{$d_1$\tablenotemark{a}}&
\colhead{$d_2$\tablenotemark{a}}&
\colhead{$d_3$\tablenotemark{a}}&
\colhead{$M_V$\tablenotemark{a}}&
\colhead{$M_I$\tablenotemark{a}}&
\colhead{$M_K$\tablenotemark{a}}
}
\startdata
0.02&0.26 & 7.50 & 7.37 & 7.28 & 271& 282& 286 & -3.51&-4.19 &-4.97  \\
0.015&0.27& 7.46 & 7.35 & 7.29 & 274& 284& 288 & -3.47&-4.16 &-4.97  \\
0.01&0.26 & 7.52 & 7.40 & 7.32 & 277& 288& 291 & -3.55&-4.23 &-5.01  \\
0.01&0.24 & 7.52 & 7.40 & 7.32 & 277& 288& 292 & -3.54&-4.21 &-5.01 
\enddata
\tablenotetext{a}{Metal ($Z$) and He ($Y$) abundance by mass; 
$V,I,K$ apparent distance moduli (mag); true distances (pc) based on the Cardelli et al. (1989) 
extinction law: $E(B-V)=(\mu_V-\mu_I)/1.22$ ($d_1$), $E(B-V)=(\mu_V-\mu_K)/2.74$ ($d_2$);  
true distances (pc) based on angular diameter variations ($d_3$); $V,I,K$ absolute magnitudines.) 
}
\end{deluxetable}
\end{center}

\clearpage
\begin{figure}
\includegraphics[height=0.55\textheight,width=0.55\textheight]{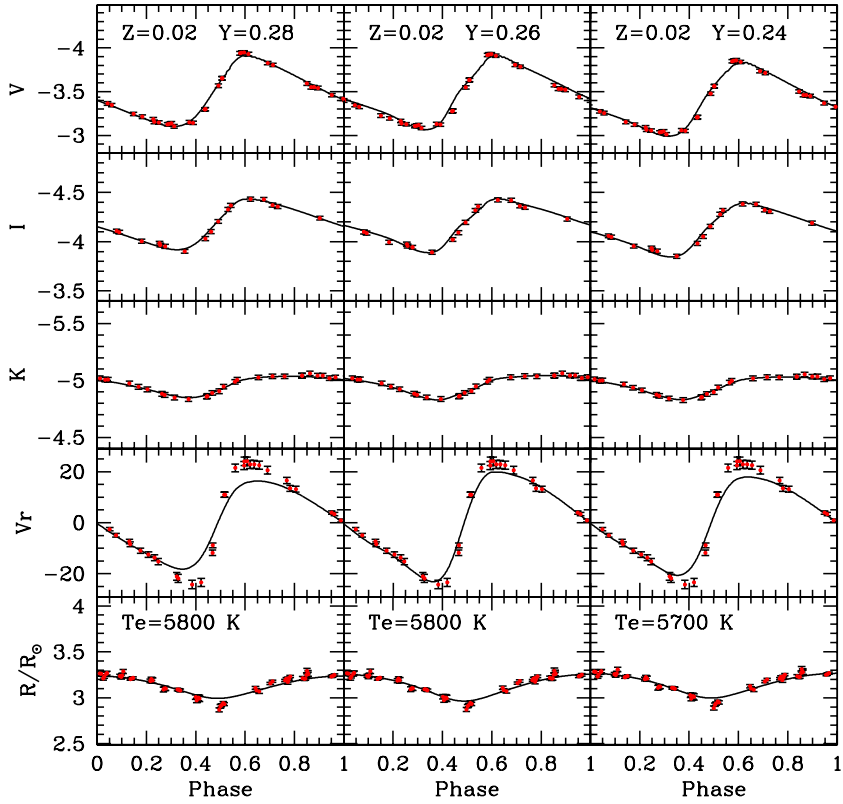}
\label{bestfit}
\caption{Comparison between best fit models at different chemical compositions
and observations. From top to bottom $V,I,K$ light curves, radial velocity 
(km/sec), and radius curve (solar units).}
\end{figure}

\clearpage
\begin{figure}
\includegraphics[height=0.55\textheight,width=0.55\textheight]{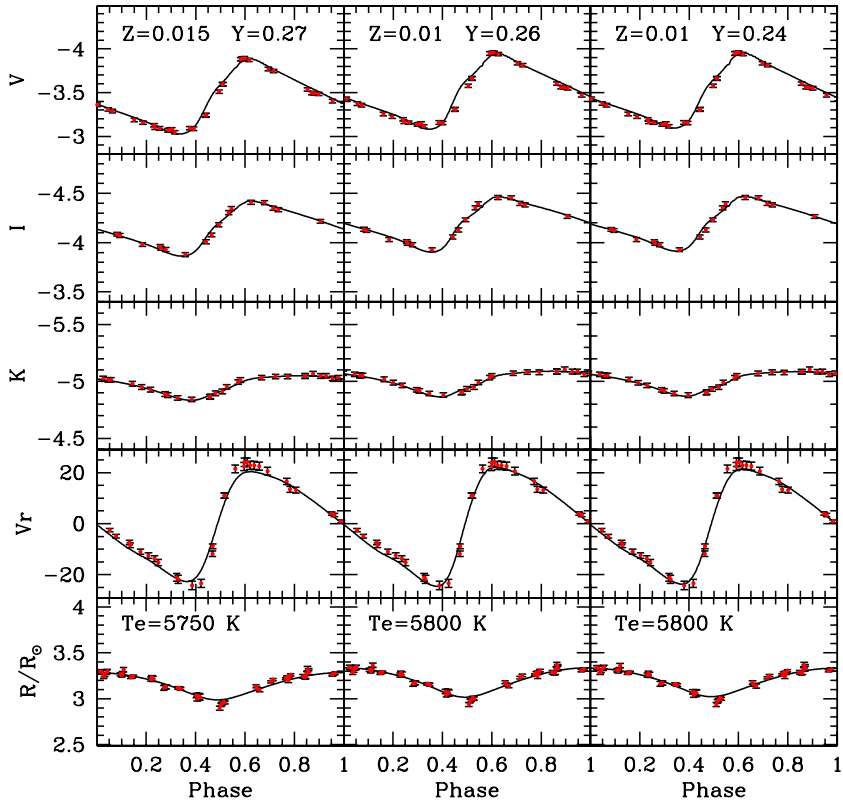}
\label{bestfit2}
\caption{Same as Fig. 1, but for more metal-poor best fit models.}
\end{figure}

\end{document}